\documentclass[9pt,twocolumn,twoside]{osajnl}

\journal{josab} 

\setboolean{shortarticle}{false} 

\title{Low-loss forward and backward surface plasmons in a~semiconductor nanowire coated by helical graphene strips}

\author[1,2]{Vitalii I. Shcherbinin}
\author[1,3]{Volodymyr I. Fesenko}
\author[1,3,*]{Vladimir R. Tuz} 

\affil[1]{International Center of Future Science, State Key Laboratory on Integrated Optoelectronics, College of Electronic Science and Engineering, Jilin University, 2699  Qianjin Str., Changchun 130012, China}
\affil[2]{National Science Center `Kharkiv Institute of Physics and Technology'of National Academy of Sciences of Ukraine, 1, Akademicheskaya Str., Kharkiv 61108, Ukraine}
\affil[3]{Institute of Radio Astronomy of National Academy of Sciences of Ukraine, 4, Mystetstv Str., Kharkiv 61002, Ukraine}
\affil[*]{Corresponding author: tvr@jlu.edu.cn; tvr@rian.kharkov.ua}

\ociscodes{(230.7370) Waveguides; (240.6680) Surface plasmons; (240.0310) Thin films; (250.5403) Plasmonics.}

\begin{abstract}
In the long-wavelength approximation, the effective conductivity tensor is introduced for graphene ribbons (strips) placed periodically at the interface between two media. The resulting conducting surface is considered as a coating for semiconductor nanowire. For the hybrid waves of such nanowire the dispersion equations are obtained in explicit form. Two types of surface plasmons are found to exist: (i)~the modified surface plasmons, which originate from the ordinary surface plasmons of a graphene-coated semiconductor nanowire, and (ii)~the spoof plasmons, which arise on the array of graphene ribbons and may possess forward-wave and backward-wave dispersion. It is revealed that the spoof surface plasmons are low-loss ones, and their frequencies, field-confinement and group velocities can be tuned widely by adjusting the coil angle and width of helical graphene strips.
\end{abstract}

\setboolean{displaycopyright}{true}

\begin{document}

\maketitle

\section{Introduction}
Graphene is known to be an actual two-dimensional material with unique electrical properties capable of being tuned by either or both chemical doping and electrical gating. Good conductivity and relatively low Ohmic losses are among its distinctive features, which make feasible propagation of surface plasmons along the graphene sheet. Owing to the intrinsic subwavelength nature, the surface plasmons can propagate in nanoscale guiding structures operating well below the diffraction limit. Such guiding, together with the strong field confinement, make surface plasmons on graphene promising candidates for use in various optoelectronic and nanophotonic devices \cite{LUO_MatSci_2015}.

From the practical standpoint, cylindrical dielectric (e.g. semiconductor) nanowires coated by graphene \cite{Gao_OE_2014, Serrano_ThzSciTech_2015, Riso_JOpt_2015, Pengchao_NP_2018} are of particular interest. Typically, such nanowires possess the core radius of several hundreds of nanometers and can support a set of surface plasmons with different azimuthal indices. Their frequencies usually do not exceed few tens of terahertz (THz) for propagation length below several micrometers. Theoretically, these frequencies can be enlarged with decrease in both the radius and permittivity of dielectric nanowire, or with increase in the chemical potential of graphene. In actual practice, such modifications, however, suffer from obvious limitations. Moreover, none of them allow for distinct tuning of the group velocities of the surface plasmons. Such a problem is topical in view of promising applications \cite{Tsakmakidis_Science_2017} of surface plasmons with nearly zero and negative group velocities. Among such applications we can mention stopped-light nanolasing \cite{Pickering_NatCommun_2014}, nanoplasmonic light-harvesting \cite{Aubry_NanoLetters_2010, Zhou_OptExpress_2017}, and on-chip spectroscopy \cite{Smolyaninova2012}. 

Ribbons (finite-width strips made of graphene), which are key elements in many graphene plasmonics studies \cite{Nikitin_PRB_2012, Tymchenko_ACSNano_2013, Abajo_AcsPhot_2014, Diaz_OptMatExpress_2015}, deserve a special attention. The quasi-one-dimensional atomic structure of graphene ribbons is similar to that of single-walled carbon nanotubes, while their electronic structures and transport are quite different. Moreover, the properties of  graphene ribbons are also highly dependent on strips size and edge shape, which can be either armchair or zigzag \cite{Nakada_PhysRevB_1996}. In the latter case, graphene ribbons twisted around a nanowire (helical graphene strips) behave as a chiral material since its left-handed and right-handed copies are given by the relative rotation of the strips. In plasmonics, both rotation directions are generally possible, thus giving rise to the electromagnetic modes rotating either clockwise or counterclockwise \cite{Kuzmin_acsphotonics_2017, Stauber_PhysRevLett_2018}. Such intrinsically chiral hybrid nanowires play a crucial role for the design of `one-way propagation' plasmonic devices with no need for an external magnetic field.

In order to describe optical properties of graphene ribbons the approach developed for dense metal-strip gratings can be applied \cite{sivov_RadioEngElectronPhys_1961, weinstein_electronics_1966, adonina_SovPhysTechPhys_1964, Bankov1988, DeLyser_JEMWA_1991, Yatsenko_AP_2000, Tretyakov_2003, Tuz_PIERB_2011}. When the period of ribbons is much smaller than the wavelength of propagating waves, the equivalent (impedance) boundary conditions can be introduced for the macroscopic fields (i.e. an array of graphene ribbons can be treated as a metasurface). These boundary conditions are obtained using the rigorous problem solution for the plane monochromatic wave diffraction on the infinite grating of metal strips placed periodically between two magneto-dielectric half-spaces. They treat the grating as an infinitely thin (semi-transparent) uniaxial anisotropic screen whose anisotropy axis is directed along the strips. 

In this paper, we consider a semiconductor nanowire coated by a metasurface composed of densely coiled helical graphene strips. We derive the dispersion equations for the waves propagating along the nanowire. It is found that, along with ordinary slow waves, such guiding structure can support spoof (designer) surface plasmons. It is our goal to show that in such a system the spoof surface plasmons exhibit low-loss propagation and may possess forward-wave and backward-wave dispersion, whereas their frequencies, field confinement and group velocities can be tuned widely by adjusting the coil angle and width of graphene strips.

\section{Effective conductivity for an array of graphene strips}
\label{sec:tensor}

Consider a metasurface in the form of two-dimensional conducting strips placed periodically at the plane interface between media \textcircled1 and \textcircled2 with the relative constitutive parameters $\varepsilon_{1}$, $\mu_{1}$ and $\varepsilon_{2}$, $\mu_{2}$, respectively. The strip width is $w$ and the period of structure is $p$. The unit vectors $\mathbf i_\parallel$ and  $\mathbf i_\perp$ directed along the principal axes of the metasurface are parallel and perpendicular to the strips, respectively.

Consider an electromagnetic wave propagating inside both media. The time-dependence of the wave field has the form $\exp(-i\omega t)$. Assume that the wave frequency $\omega$ satisfies the condition $p/\lambda = N/2 \ll 1$, where $\lambda=2\pi/k$ is the wavelength, and $k$ is the wavenumber in free space. In this long-wavelength approximation, there is a negligible contribution of the higher Floquet harmonics to the wave field at the metasurface area, where the averaged boundary conditions can be used. For the metasurface composed of the perfectly conducting strips, these conditions have the following form \cite{Bankov1988}:

\begin{figure}[tbp]
\centering
\includegraphics[width=0.6\linewidth]{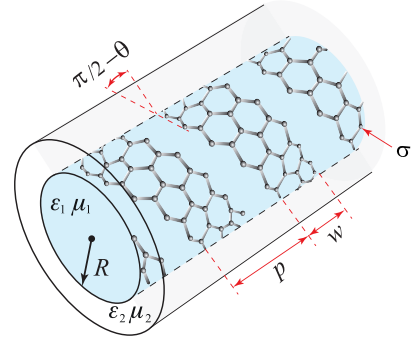}
\caption{Cylindrical semiconductor nanowire (blue colored region) coated by helical graphene strips.}
\label{fig:fig1}
\end{figure}

\begin{equation}
\begin{split}
& E_\parallel^\textrm{\textcircled1} = E_\parallel^\textrm{\textcircled2},~ H_\parallel^\textrm{\textcircled2} - H_\parallel^\textrm{\textcircled1} = -Z_\perp^{-1}E_\perp^\textrm{\textcircled1}, \\
& E_\perp^\textrm{\textcircled1} = E_\perp^\textrm{\textcircled2},~ Z_\parallel\left(H_\perp^\textrm{\textcircled2} - H_\perp^\textrm{\textcircled1}\right) = E_\parallel^\textrm{\textcircled1},
\label{eq:boundary} 
\end{split}
\end{equation}
where $Z_\perp^{-1} = -iZ^{-1}N\varepsilon_{\alpha}q\ln[\sin^{-1}(0.5\pi(1- u))]$, $Z_\parallel = -iZN\mu_{\alpha}q\ln[\sin^{-1}(0.5\pi u)]$, $\varepsilon_{\alpha} = \varepsilon_{1} + \varepsilon_{2}$, $\mu_{\alpha}=\mu_{1}\mu_{2}/(\mu_{1} + \mu_{2})$, $u=w/p$, $q = 1 -k_\parallel^2/(\varepsilon_{\alpha}\mu_{\alpha} k^2)$, $k_\parallel^2=-\partial^2/\partial x_\parallel^2$, $x_\parallel$ is the coordinate along the strips, and $Z$ is the impedance of free space.

The averaged boundary conditions (\ref{eq:boundary}) are at least first-order in $p/\lambda$ \cite{DeLyser_JEMWA_1991, Yatsenko_AP_2000, Tretyakov_2003} and are valid for waves propagating in an arbitrary direction with respect to the axis of the strips. Note that $Z_\parallel$ ($Z_\perp$) takes the well-known form \cite{Tretyakov_2003, Luukonen_AP_2008} of the surface impedance for the non-magnetic ($\mu_{1} = \mu_{2} = 1$) planar metasurface excited by the normally incident plane wave whose electric field vector is directed parallel (perpendicular) to the strips.

As $u \to 0$, both $Z_\parallel$ and $Z_\perp$ approach infinity and \eqref{eq:boundary} reduces to the field continuity conditions at the interface between two media. The opposite extreme case is $u \to 1$. In this case both $Z_\parallel$ and $Z_\perp$ vanish and the metasurface becomes a perfect electric conductor.

Averaged boundary conditions (\ref{eq:boundary}) can be extended \cite{Fallahi_APL_2012} to the metasurface made of graphene strips with the finite conductivity $\sigma(\omega)$. For this purpose, we introduce the graphene surface impedance $Z_g^{av} = u\sigma^{-1}$ averaged over the metasurface area. With this impedance in hand, we can rewrite the boundary conditions (\ref{eq:boundary}) at the metasurface area as

\begin{equation}
\begin{split}
& E_\parallel^\textrm{\textcircled1} = E_\parallel^\textrm{\textcircled2},~ H_\parallel^\textrm{\textcircled2} - H_\parallel^\textrm{\textcircled1} = -\sigma_\perp E_\perp^\textrm{\textcircled1}, \\
& E_\perp^\textrm{\textcircled1} = E_\perp^\textrm{\textcircled2},~ H_\perp^\textrm{\textcircled2} - H_\perp^\textrm{\textcircled1} =\sigma_\parallel E_\parallel^\textrm{\textcircled1},
\label{eq:boundary_gph} 
\end{split}
\end{equation}
where $\sigma_{\perp,\parallel} = (Z_{\perp,\parallel} + Z_g^{av})^{-1}$ are the principal components of the metasurface conductivity tensor
\begin{equation}
\hat {\boldsymbol \sigma} =\left( {\begin{matrix}
   {\sigma_\perp} & {0} \cr
   {0} & {\sigma_\parallel} \cr
\end{matrix}
} \right). \label{eq:sigma}
\end{equation}

Note that the component $\sigma_\perp$ of the effective conductivity tensor (\ref{eq:sigma}) coincides with that obtained for a hyperbolic graphene-based metasurface 
when $\mu_{1} = \mu_{2} = 1$, $\varepsilon_{1} = \varepsilon_{2} = 1$, and $\partial/\partial x_\parallel = 0$ (see \cite{Diaz_OptMatExpress_2015}). Moreover, the component $\sigma_\parallel$ has the same form as for a grid of lossy wires with the averaged surface impedance $Z_g^{av}=pZ_w$ \cite{Tretyakov_2003}, where $Z_w$ is the wire impedance per unit length.

Let us introduce an additional pair of orthogonal unit vectors $\mathbf i_z$ and $\mathbf i_\varphi$ rotated by the angle $\theta$ with respect to the pair $\mathbf i_\parallel$ and $\mathbf i_\perp$. Generally, in the coordinates $\{\varphi,z\}$, the conductivity tensor (\ref{eq:sigma}) is non-diagonal and the boundary conditions (\ref{eq:boundary_gph}) take the form \cite{Yermakov_PhysRevB_2015}:
\begin{equation}
\begin{split}
& E_z^\textrm{\textcircled1} = E_z^\textrm{\textcircled2},~ H_z^\textrm{\textcircled2} - H_z^\textrm{\textcircled1} = -\sigma_{\varphi\varphi} E_\varphi^\textrm{\textcircled1} - \sigma_{\varphi z} E_z^\textrm{\textcircled1}, \\
& E_\varphi^\textrm{\textcircled1} = E_\varphi^\textrm{\textcircled2},~ H_\varphi^\textrm{\textcircled2} - H_\varphi^\textrm{\textcircled1} =\sigma_{zz} E_z^\textrm{\textcircled1} + \sigma_{z\varphi} E_\varphi^\textrm{\textcircled1},
\label{eq:boundary_tens} 
\end{split}
\end{equation}
where $\sigma_{\varphi\varphi} = \sigma_\perp\cos^2\theta + \sigma_\parallel \sin^2\theta$, $\sigma_{zz} = \sigma_\parallel \cos^2\theta + \sigma_\perp\sin^2\theta$, and $\sigma_{\varphi z} = \sigma_{z\varphi} = (\sigma_\parallel - \sigma_\perp)\sin\theta\cos\theta$.

Assume for the moment that the electromagnetic wave under consideration has no variations along the metasurface area and thus $\partial/\partial x_\parallel = 0$. In such a case we can distinguish two types of waves satisfying \eqref{eq:boundary_tens} for $\theta=0^\circ$ and $\theta = 90^\circ$ ($\sigma_{\varphi z} = \sigma_{z\varphi} = 0$). For the waves of one type, $E_\varphi = 0$  ($H_z =0$). When $\theta = 0^\circ$ ($\theta = 90^\circ$), their electric field $E_z$ is parallel (perpendicular) to the graphene strips and, in line with \cite{Tretyakov_2003, Luukonen_AP_2008}, $\sigma_{zz}$ reduces to $\sigma_\parallel$ ($\sigma_\perp$). Evidently, the electromagnetic properties of these waves are independent of $\sigma_\perp$ ($\sigma_\parallel$). By contrast, for the waves of another type, $E_z = 0$ ($H_\varphi = 0$) regardless of $\sigma_{zz}$. When $\theta = 0^\circ$ or $\theta = 90^\circ$, their electric field $E_\varphi$ is polarized either perpendicular or parallel to the graphene strips, respectively. As might be expected \cite{Tretyakov_2003, Luukonen_AP_2008}, in this case $\sigma_{\varphi\varphi}$  is equal to $\sigma_\perp$ or $\sigma_\parallel$.

Note that the expressions (\ref{eq:boundary}) related to $Z_\parallel$ and $Z_\perp$ were derived in \cite{Bankov1988, DeLyser_JEMWA_1991, Yatsenko_AP_2000, Tretyakov_2003, Luukonen_AP_2008} under an assumption that the interface between media \textcircled1 and \textcircled2 is planar. However, as was shown in \cite{Sipus_MMET_1998, Padooru_JAP_2012}, these expressions may be successfully used to study cylindrical metasurfaces, while being highly accurate for $N \le 0.25$. In what follows, the value $N = 0.05/\pi$ is used for all our calculations.

\section{Independent dispersion equations for TE-like and TM-like waves}

Let cylindrical graphene-based metasurface of radius $R$ be the interface between the core and cladding regions filled by media \textcircled1 and \textcircled2, respectively (Fig.~\ref{fig:fig1}). Of interest are the waves propagating along the structure axis parallel to $\mathbf i_z$. The wave fields have the following form: $\{\mathbf{E}(\mathbf{r}),\mathbf{H}(\mathbf{r})\}\exp(-i\omega t+ik_zz+il\varphi)$, where $k_z$ and $l$ are the axial and the azimuth wavenumbers, respectively. In this case $\partial^2/\partial x_\parallel^2 = -k_\parallel^2 = -(k_z\cos\theta+lR^{-1}\sin\theta)^2$.

The standard scheme of deducing the dispersion equation is as follows. From Maxwell's equations, the wave fields are expressed in terms of cylindrical [Bessel $J_l(\cdot)$ and Neumann $N_l(\cdot)$] functions with unknown integration constants. The number of unknown constants equals the number of independent boundary conditions. The resultant system of equations has a nontrivial solution if the determinant of its matrix is zero. From this requirement follows the dispersion equation for the wave frequency $\omega(k_z)$. Evidently, four independent boundary conditions (\ref{eq:boundary_tens}) must lead to the dispersion equation in the form of vanishing determinant of a $4 \times 4$ matrix \cite{Gao_OE_2014, Pengchao_NP_2018}. In fact, this equation can be greatly simplified.

To demonstrate this, we follow the procedure outlined in \cite{Shcherbinin_PIERM_2017} and first reduce \eqref{eq:boundary_tens} to two one-side boundary conditions imposed on the field of the core region at $r = R$
\begin{equation}
\left( {\begin{matrix}
   {H_z^\textrm{\textcircled1}} \cr
   {-H_\varphi^\textrm{\textcircled1}} \cr
\end{matrix}
} \right) = \mathbf{\hat Y} \left( {\begin{matrix}
   {E_\varphi^\textrm{\textcircled1}} \cr
   {E_z^\textrm{\textcircled1}} \cr
\end{matrix}
} \right) = \left( {\begin{matrix}
   {Y_{\varphi\varphi}} & {Y_{\varphi z}} \cr
   {Y_{z\varphi}} & {Y_{zz}} \cr
\end{matrix}
} \right) \left( {\begin{matrix}
   {E_\varphi^\textrm{\textcircled1}} \cr
   {E_z^\textrm{\textcircled1}} \cr
\end{matrix}
} \right), \label{eq:admittance}
\end{equation}
where $\mathbf{\hat Y}$ is the tensor of the metasurface admittance with the components $Y_{\varphi\varphi} = \eta_\varphi^{-1} + \sigma_{\varphi\varphi}$, $Y_{\varphi z} = Y_{z\varphi} = \sigma_{\varphi z} + \eta_\varphi^{-1}lk_z/(\mbox{\ae}_2^2 R)$, $Y_{zz} = Y_{z\varphi} = \eta_z^{-1} + \sigma_{zz} + \eta_\varphi^{-1}[lk_z/(\mbox{\ae}_2^2 R)]^2$, and $\mbox{\ae}_2^2 = \varepsilon_{2}\mu_{2}k^2-k_z^2$.

Note that the one-side boundary conditions (\ref{eq:admittance}) at the metasurface area have the same form as those used to describe tensor impedance surfaces \cite{Patel_IEEE-MTT_2013,Quarfoth_IEEE-AP_2013,Patel_IEEE-MTT_2014} designed for X-band of microwave spectrum.

In \eqref{eq:admittance}, the fields in the cladding region are introduced by two functions
\begin{equation}
\begin{split}
& \eta_\varphi = -iZ\frac{k\mu_{2}}{\mbox{\ae}_2^2} \frac{1}{H_z^\textrm{\textcircled2}} \left.\frac{dH_z^\textrm{\textcircled2}}{dr}\right|_{r=R}, \\
& \eta_z^{-1} = -iZ^{-1}\frac{k\varepsilon_{2}}{\mbox{\ae}_2^2} \frac{1}{E_z^\textrm{\textcircled2}} \left.\frac{dE_z^\textrm{\textcircled2}}{dr}\right|_{r=R},
\label{eq:cladding} 
\end{split}
\end{equation}
which can be determined in an explicit form for several waveguide configurations. Among them there is a circular step-index waveguide with an unbounded cladding region \textcircled2. In this case one can write

\begin{equation}
H_z^\textrm{\textcircled2} = A_2 H_l^{(1)}(\mbox{\ae}_2r),~~E_z^\textrm{\textcircled2} = B_2 H_l^{(1)}(\mbox{\ae}_2r), 
\label{eq:fieldsin2} 
\end{equation}

\begin{equation}
\eta_\varphi = -iZ\frac{k\mu_{2}}{\mbox{\ae}_2} \frac{H_l^{'(1)}(\mbox{\ae}_2R)}{H_l^{(1)}(\mbox{\ae}_2R)},~~\eta_z^{-1} = -iZ^{-1}\frac{k\varepsilon_{2}}{\mbox{\ae}_2} \frac{H_l^{'(1)}(\mbox{\ae}_2R)}{H_l^{(1)}(\mbox{\ae}_2R)}, 
\label{eq:admitancein2} 
\end{equation}
where $H_l^{(1)}(\cdot)=J_l(\cdot)+iN_l(\cdot)$ is the Hankel function of the first kind, $A_2$ and $B_2$ are the unknown field amplitudes, which have no effect on $\eta_\varphi$ and $\eta_z$.

In view of \eqref{eq:admitancein2}, the metasurface admittance $\mathbf{\hat Y}$ in \eqref{eq:admittance} is a function of the desired wave frequency $\omega(k_z)$ and given parameters, such as the azimuth wavenumber $l$, the radius $R$, the permittivity $\varepsilon_{2}$, the permeability $\mu_{2}$, the effective metasurface conductivity (\ref{eq:sigma}), and the angle $\theta$. The one-side boundary conditions (\ref{eq:admittance}) at the metasurface $r = R$ can be rewritten as:
\begin{equation}
\begin{split}
& \frac{dH_z^\textrm{\textcircled1}}{dr}+aH_z^\textrm{\textcircled1}+ibZ^{-1}\sqrt{\frac{\varepsilon_{1}}{\mu_{1}}}E_z^\textrm{\textcircled1}=0, \\
&  \frac{dE_z^\textrm{\textcircled1}}{dr}+cE_z^\textrm{\textcircled1}-ibZ\sqrt{\frac{\mu_{1}}{\varepsilon_{1}}}H_z^\textrm{\textcircled1}=0,
\label{eq:Eq9} 
\end{split}
\end{equation}
where $b=-(lk_z -R\mbox{\ae}_1^2y_{\varphi z}y^{-1}_{\varphi\varphi})/ (kR\sqrt{\varepsilon_{1}\mu_{1}})$, $c = -i( y_{zz}-y^2_{\varphi z}y^{-1}_{\varphi\varphi})\mbox{\ae}_1^2 / (k\varepsilon_{1})$, $a=-iy^{-1}_{\varphi\varphi}\mbox{\ae}_1^2/(k\mu_{1})$, $\mbox{\ae}_1^2 = \varepsilon_{1}\mu_{1}k^2 - k_z^2$, and $\mathbf{\hat y}=Z\mathbf{\hat Y}$ is the dimensionless admittance tensor. As both $\sigma_\perp$ and $\sigma_\parallel$ approach zero, \eqref{eq:Eq9} reduces to the results obtained in \cite{Shcherbinin_PIERM_2017}.

Generally, substitution of the core fields $\left\lbrace E_z^\textrm{\textcircled1}(r), H_z^\textrm{\textcircled1}(r)\right\rbrace$ into \eqref{eq:Eq9} leads to the dispersion equation in the form of vanishing determinant of a $2\times2$ matrix. But further simplification can be obtained for the homogeneous isotropic core under consideration.

The reason is that in this case both $E_z^\textrm{\textcircled1}(r)$ and $H_z^\textrm{\textcircled1}(r)$ are proportional to the same membrane function $\psi(r)=A_1J_l(\mbox{\ae}_1r)$, where $A_1$ is an arbitrary field amplitude. Therefore one can write: $H_z^\textrm{\textcircled1}(r)=\psi(r)$ and $E_{z1}(r)=-iPZ\sqrt{\mu_{1}/\varepsilon_{1}}\psi(r)$, where $P$ is the hybridization parameter. As a result, from \eqref{eq:cladding} follow two conditions on $\psi(r)$ which yield two independent dispersion equations \cite{Shcherbinin_PIERM_2017,Shcherbinin_PAST_2015}
\begin{equation}
\begin{split}
& \mbox{\ae}_1 J'_l\left(\mbox{\ae}_1R\right) + g_1J_l\left(\mbox{\ae}_1R\right) = 0, \\
& \mbox{\ae}_1 J'_l\left(\mbox{\ae}_1R\right) + g_2J_l\left(\mbox{\ae}_1R\right) =0,
\label{eq:DispEq} 
\end{split}
\end{equation}
where $g_{1,2}=(a+bP_{1,2})=(c+bP^{-1}_{1,2})$, $P_{1,2}=\alpha\left(1 \mp\sqrt{1+\alpha^{-2}} \right)$ are two values of the hybridization parameter $P$, $\alpha=(c-a)/2b$, $P_2=-1/P_1$.

It is easy to show \cite{Shcherbinin_PAST_2015} that $P_1$ and $P_2$ always satisfy the conditions $|P_1| \leq 1$ and $|P_2| \geq 1$. Thus the upper (lower) equation of Eq.~(\ref{eq:DispEq}) describes hybrid TE-like (TM-like) waves of the step-index waveguide containing a graphene-based metasurface. These waves become pure TE ($P=P_1=0$) and TM ($P=P_2=\infty$) modes as $\alpha$ approaches infinity (e.g. $b \to 0$). An example is the axially symmetric ($l=0$) modes of the step-index waveguide in the case of $\theta=0^\circ$ (axial graphene strips) or $\theta=90^\circ$ (azimuthal graphene strips).

\begin{figure}[htbp]
\centering
\includegraphics[width=0.8\linewidth]{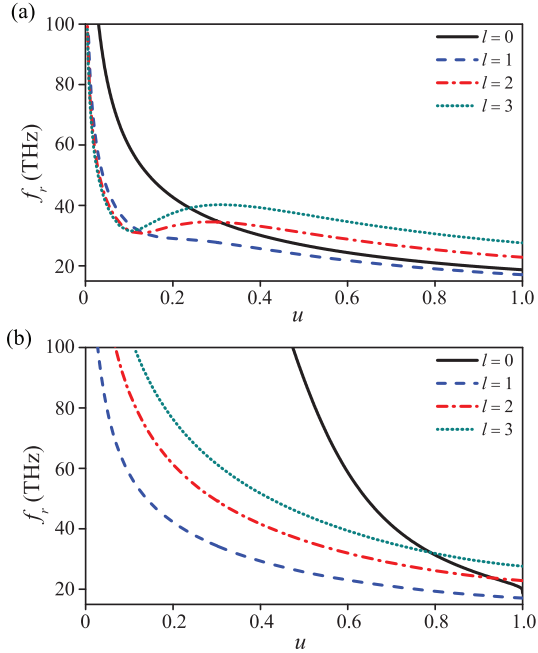}
\caption{Frequencies of the modified surface plasmons (MSPs) of semiconductor nanowire coated with graphen strips versus $u$; (a) axial strips ($\theta = 0^\circ$); (b) azimuthal strips ($\theta = 90^\circ$); $k_zR=0.5$.}
\label{fig:fig2}
\end{figure}

The independent dispersion equations (\ref{eq:DispEq}) of TE-like and TM-like waves are of the similar form, which is relatively easy to analyze. If $g_1$ and $g_2$ are real (lossless case) and satisfy the conditions $g_1R+|l|<0$ and $g_2R+|l|<0$, these equations yield purely imaginary eigenvalues $\chi=\mbox{\ae}_1R$ \cite{watson1995treatise}, which correspond to slow waves (surface plasmons). In what follows, our prime interest is in the TM-like ($P=P_2$) surface plasmons.

\section{Modified surface plasmons and spoof plasmons}

In numerical calculations we consider a graphene-coated semiconductor nanowire described in \cite{Pengchao_NP_2018}. The nanowire parameters are as follows: $\mu_{1}=\mu_{2}=1$, $R=50$~nm, $\varepsilon_1 =12.25$, $\varepsilon_2 = 1$. The graphene conductivity $\sigma = \sigma(\omega)$ (see Fig.~1(b) of \cite{Pengchao_NP_2018}) is found from the Kubo formula \cite{Falkovsky_PhysRev_2007} for the temperature $T=300$~K, the chemical potential $\mu_{ch}=0.5$~eV, and the relaxation time $\tau=1.84 \times 10^{13}$~s. First we ignore the Ohmic losses and thus set Re$(\sigma)=0$. As was found in \cite{Pengchao_NP_2018}, in the case of homogeneous graphene coating ($u=1$) the nanowire in hand can support several TM-like surface plasmons with different azimuth indices $l$. As $u$ decreases, their frequencies $f_r=\omega/2\pi$ generally increase. This is shown in Fig.~\ref{fig:fig2}(a) and Fig.~\ref{fig:fig2}(b) for the axial and azimuthal graphene strips, respectively. Similar increase in plasmon frequency with increasing gaps between graphene strips was observed in \cite{Fallahi_APL_2012,Nikitin_PRB_2012,Shapoval_IEEE-TS_2013}. This lends support to the validity of our results. As noted in \cite{Fallahi_APL_2012,Tymchenko_ACSNano_2013}, such frequency change can be attributed to the fact that the periodic gaps introduced in the graphene sheet enhance the effective capacitance of the metasurface and thereby increase the imaginary part of its effective conductivity.

\begin{figure}[tbp]
\centering
\includegraphics[width=0.8\linewidth]{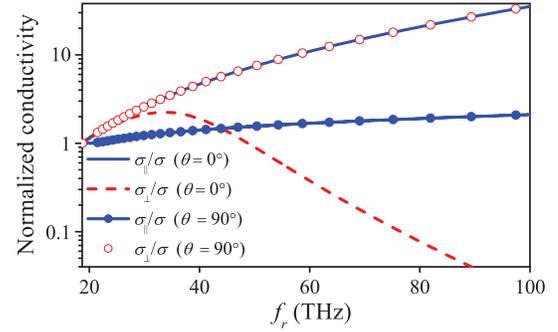}
\caption{Normalized components of the metasurface conductivity versus the frequency $f_r(u)$ shown in Fig.~\ref{fig:fig2} for the axially symmetric surface plasmon ($l=0$).}
\label{fig:fig3}
\end{figure}

This can be demonstrated by the example of the axially symmetric surface plasmon. As discussed above, for $\theta= 0^\circ$ and $\theta= 90^\circ$ the electromagnetic properties of such plasmon are independent of $\sigma_\perp$ and $\sigma_\parallel$, respectively. Fig.~\ref{fig:fig3} shows the change of the metasurface conductivity along the curves $f_r(u)$ depicted in Fig.~\ref{fig:fig2} for the axially symmetric TM plasmon at $k_z R= 0.5$ . As can be seen from Figs.~\ref{fig:fig2} and \ref{fig:fig3}, the smaller is $u$, the higher are both the metasurface conductivity and the plasmon frequency. The practical outcome of this is that the filling factor $u$ can be used to tune the frequencies of surface plasmons in a wide range. Such property of plasmons relates them to spoof (designer) counterparts. Despite this, we will call them modified surface plasmons (MSPs), since they originate from the ordinary surface plasmons of a graphene-coated nanowire \cite{Pengchao_NP_2018}.

Numerical calculations reveal another spoof plasmons of the nanowire in hand. No such waves exist in the original guiding structure with the homogeneous graphene coating. The spoof plasmons are formed as soon as the gaps appear between the graphene strips.

The frequencies of these plasmons are generally close to solutions of the following equation
\begin{equation}
\sigma^{-1}_\perp\left(\omega,k_z \right)=Z_{\perp}+Z_g^{av}=0.
\label{eq:SpoofPlasmons}
\end{equation}

\begin{figure}[tbp]
\centering
\includegraphics[width=0.8\linewidth]{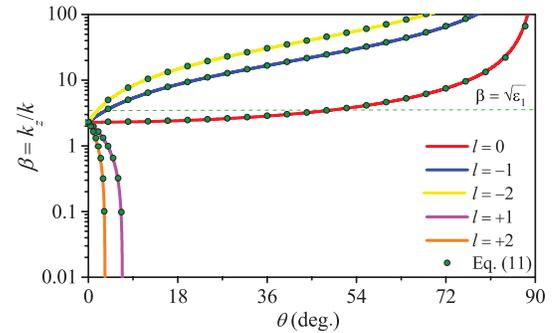}
\caption{Normalized propagation constant as a function of $\theta$ for spoof plasmons with different azimuth indices $l$; $f_r=50$~THz, $u=0.9$.}
\label{fig:fig4}
\end{figure}

It is clear that in this case the metasurface behaves as a good conductor in the direction perpendicular to the graphene strips. For small gaps ($1 - u \ll 1$) and high frequencies ($Z_g^{av}\gg Z$), \eqref{eq:SpoofPlasmons} can be approximated as
\begin{equation}
\varepsilon_{\alpha}\mu_{\alpha}k^2-k^2_\parallel=\frac{1}{2}\left(\varepsilon_{1}+\varepsilon_{2}\right) k^2-k^2_\parallel=0.
\label{eq:SpoofPlasmons_Approx}
\end{equation}

\begin{figure}[tbp]
\centering
\includegraphics[width=0.8\linewidth]{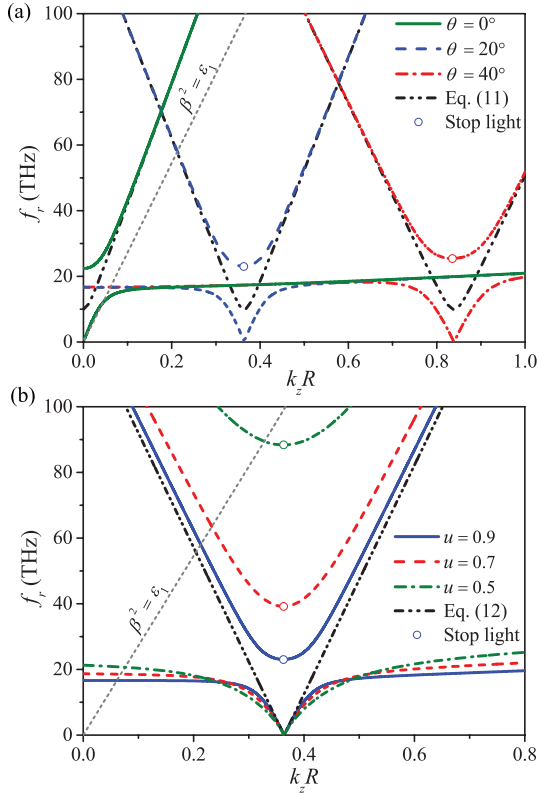}
\caption{Dispersion curves of spoof plasmon ($l=-1$) for: (a) $u=0.9$ and different $\theta$; (b) $\theta=20^\circ$ and different $u$.}
\label{fig:fig5}
\end{figure}

It describes the wave propagating along the strips with the phase velocity equal to $\omega/k_\parallel = c\sqrt{2/\left(\varepsilon_{1} + \varepsilon_{2}\right)}$. Evidently, this wave is fast ($k_z < \sqrt{\varepsilon_{1}}k$) in the case of semiconductor nanowire coated by the axial graphene strips ($\theta = 0^\circ$, $k_\parallel = k_z$).

The observed spoof plasmons resemble the so-called `grating' \cite{Bellamine_MTT_1994} or `strip' \cite{Sipus_IEEProc_1997} modes, which were found to exist in a plane grounded slab loaded with strips. According to \cite{Sipus_IEEProc_1997}, such modes propagate along the strips with the wavevector $k_\parallel$ defined by \eqref{eq:SpoofPlasmons_Approx}. Besides, they are reported \cite{Bellamine_MTT_1994} to become slow wave modes (surface plasmon) as $\theta$ deviates from zero and to be highly localized surface waves as $\theta$ tends to $90^\circ$. For $l\leq 0$ the last property of spoof plasmons of the semiconductor nanowire coated by graphene strips ($u =0.9$) can be seen from Fig.~\ref{fig:fig4}, which shows the normalized propagation constant $\beta=k_z/k$ versus $\theta$ for plasmons with the frequency of 50~THz. For comparison, this figure also depicts solutions of \eqref{eq:SpoofPlasmons}, which are seen to be very close to the frequencies $\omega(k_z)$ of spoof plasmons regardless of $\theta$. Fig.~\ref{fig:fig4} is similar to Fig.~2 of \cite{Bellamine_MTT_1994} for the `grating' mode of a strip loaded grounded slab. Although the result of the latter figure follows from the approximate homogenization analysis of \cite{DeLyser_JEMWA_1991}, it was later validated against the full wave calculations (see Fig.~5 in \cite{Baccarelli_IEEE_AP_2005}). It can be seen from Fig.~\ref{fig:fig4} that the phase velocities are very different for the right-handed ($l>0$) and the left-handed ($l< 0$) circularly polarized spoof plasmons. As might be expected, with the sign change of $\theta$ the right-polarized plasmons behave like left-polarized ones and vice versa. Such nonreciprocity effect is associated with chirality of the helical graphene strip and can be used for non-magnetic optical rotation of linearly-polarized waves propagating along the semiconductor nanowire. Although this effect is of practical importance, it is beyond the scope of our present study.

\section{Anti-crossing effect}

Fig.~\ref{fig:fig4} suggests that for $\theta>0$ the right-handed ($l>0$) circularly polarized spoof plasmon are fast (bulk) plasmons. For this reason they are not able to interact with slow-wave MSPs. In contrast, the left-polarized spoof plasmons can be slow waves in this case. Therefore, the possibility exists of their interaction with the modified surface plasmons.

Such is indeed the case. The interaction between spoof plasmons and MSP can be seen from Fig.~\ref{fig:fig5}(a) for $l=-1$. For high frequencies the dispersion curves of spoof plasmons are close to solutions of \eqref{eq:SpoofPlasmons}, which thus can be considered as the frequencies $\omega(k_z)$ of uncoupled spoof plasmons. These solutions may intersect the dispersion curve of MSP. However, in actual conditions there are no such intersections due to coupling between the spoof surface plasmons and the modified surface plasmon. In this case the clear-cut anti-crossing effect \cite{Tuz_Superlattices_2017} is observed.

As a result, for any $\theta>0^\circ$ there are high-frequency and low-frequency surface plasmons (Fig.~\ref{fig:fig5}), which originate from the spoof plasmon and MSP, respectively. Each of them features forward-wave and backward-wave branches of the dispersion curve. The group velocities of these surface plasmons changes rapidly near $k_z R=-l\tan \theta$ $(k_\parallel=0)$. At such axial wavenumber, the high-frequency surface plasmon possesses zero group velocity (stopped light point), while the low-frequency surface plasmon exhibits zero frequency. As $u$ decreases, the high-frequency surface plasmon shifts to higher frequencies [Fig.~\ref{fig:fig5}(b)] and finally becomes bulky. Thus with $u$ and $\theta$ one can tune the stopped light point of the surface plasmon in a wide region of the $\omega-k_z$ plane and thereby attain the desired values for both plasmon frequency and field confinement. It is interesting to note that the dispersion curves of the high-frequency and the low-frequency surface plasmons [Fig.~\ref{fig:fig5}(b)] are somewhat similar to those of upper and lower polaritons \cite{Piccione_RPP_2014}. In this case the pure `strip' wave of Eq.~(\ref{eq:SpoofPlasmons_Approx}) plays the role of `photon' in `photon-plasmon' coupling.

\section{Effect of losses}

In this section, allowance is made for interband and intraband losses in graphene, while semiconductor losses are still neglected because of their minor effect on surface plasmons \cite{Lovat_IEEE_EC_2013, Fuscaldo_JIMTW_2015}. Obviously, nonzero losses limit the propagation length $L_{pr}=1/|\text{Im}(k_z)|$ for the plasmons of semiconductor nanowire and thereby can negate the above-described beneficial effects of graphene-based coating. Therefore, we examine these effects with regard to graphene losses.

\begin{figure}[tbp]
\centering
\includegraphics[width=0.8\linewidth]{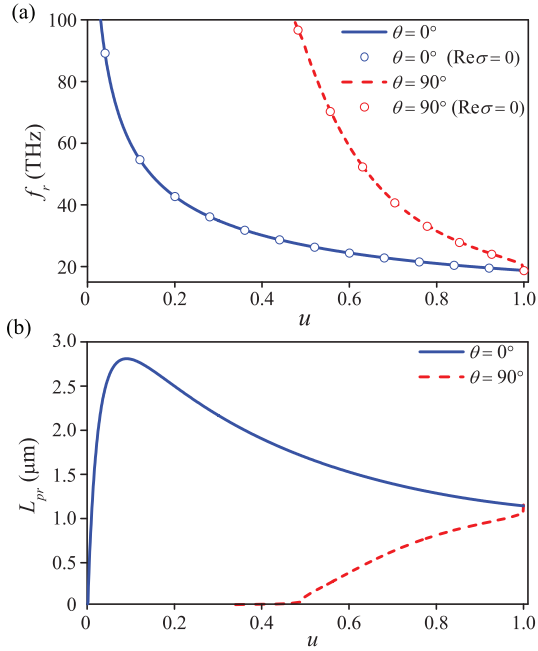}
\caption{(a) Frequency and (b) propagation length versus graphene filling factor $u$ for the axially symmetric MSP at $k_zR=0.5$.}
\label{fig:fig6}
\end{figure}

\begin{figure}[tbp]
\centering
\includegraphics[width=0.8\linewidth]{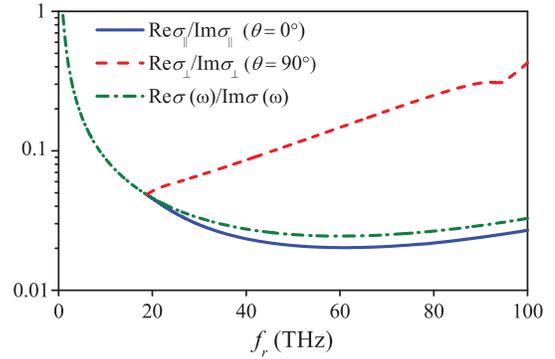}
\caption{$\text{Re}(\sigma_\parallel) / \text{Im}(\sigma_\parallel)$ and $\text{Re}(\sigma_\perp) / \text{Im}(\sigma_\perp)$ versus the frequency $f_r(u)$ shown in Fig.~\ref{fig:fig6}(a) for the axially symmetric surface plasmon ($l=0$) in the cases of $\theta = 0^\circ$ and $\theta =90^\circ$, respectively; the dependence of $\text{Re}(\sigma) / \text{Im}(\sigma)$  on the frequency is given for reference; $k_zR=0.5$.}
\label{fig:fig7}
\end{figure}

First we study the possibility of wide-band variation of MSP’s frequency with $u$. Calculations reveal that graphene losses have only slight effect on this frequency regardless of $u$ [Fig.~\ref{fig:fig6}(a)], while the MSP’s propagation length may change considerably [Fig.~\ref{fig:fig6}(b)]. As Fig.~\ref{fig:fig6}(b) suggests, for $\theta=0^\circ$ and $\theta=90^\circ$ the propagation length of MSP shows opposite changes with $u$. This is due to the opposite effects of graphene losses on the components $\sigma_\perp$ and $\sigma_\parallel$ of the metasurface conductivity. Fig.~\ref{fig:fig7} shows $\text{Re}(\sigma_\parallel) / \text{Im}(\sigma_\parallel)$ and $\text{Re}(\sigma_\perp) / \text{Im}(\sigma_\perp)$ against the frequency of the axially symmetric MSP in the case of $\theta=0^\circ$ and $\theta=90^\circ$, respectively. Here $\text{Im}(\sigma_\parallel)=\text{Im}(\sigma_\perp)$ (Fig.~\ref{fig:fig3}). It can be seen that in the high-frequency band the value of $\text{Re}(\sigma_\perp) / \text{Im}(\sigma_\perp)$ for the metasurface made of the azimuthal graphene strips ($\theta=90^\circ$) is distinctly higher as opposed to graphene loss tangent $\text{Re}[\sigma(\omega)] / \text{Im}[\sigma(\omega)]$. Therefore, although there is a rapid increase in plasmon frequency with $u$ in this case [Fig.~\ref{fig:fig6}(a)], it is associated with extremely high attenuation of MSP [Fig.~\ref{fig:fig6}(b)]. It is interesting to note that for incident light polarized perpendicular to the graphene strips a distinct absorption peak originating from plasmon oscillation was experimentally observed in \cite{Ju_NatNano_2011}. This may serve as an additional argument in favour of our description.

The opposite situation occurs with the nanowire coating composed of the axial ($\theta=0^\circ$) graphene strips. Compared to homogeneous graphene, such coating features lower losses (Fig.~\ref{fig:fig7}). For this reason, the propagation length  $L_{pr}$ of MSP enhances with increasing axial gaps cut through graphene coating. From Fig.~\ref{fig:fig6} follows that such gaps provide a way for several-fold increase in frequency of plasmon without reducing its propagation length. Note that for the semiconductor nanowire with uniform graphene coating \cite{Pengchao_NP_2018} it is hardly feasible to attain such frequency enhancement in the THz band due to limitations on the nanowire radius, core permittivity and graphene conductivity. 

\begin{figure}[tbp]
\centering
\includegraphics[width=0.8\linewidth]{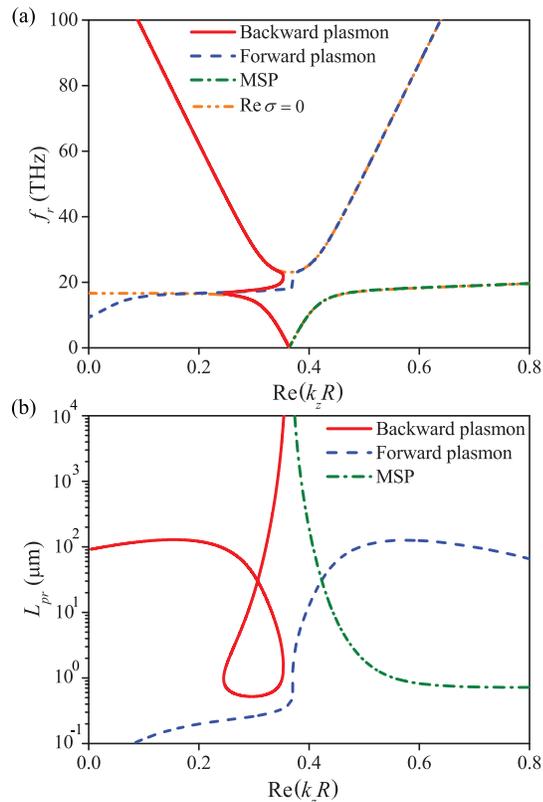}
\caption{(a) Frequency and (b) propagation length of plasmons versus Re$(k_zR)$; $l=-1$, $u=0.9$, $\theta=20^\circ$.}
\label{fig:fig8}
\end{figure}

Next we investigate the effect of graphene losses on the dispersion curves of high-frequency and low-frequency surface plasmons of Fig.~\ref{fig:fig5}. The dispersion curves are found to split into three branches [Fig.~\ref{fig:fig8}(a)]. Unfortunately, none of these branches possesses actual stopped light points due to losses. The first branch is the low-frequency forward-wave MSP with no cutoff frequency. As Re$(k_z)$ approaches $(-l/R)\tan(\theta)$ from the right, the frequency of this branch tends to zero, together with losses of MSP (Fig.~\ref{fig:fig8}).

Other branches are the forward and the backward plasmons, which both cover a broad frequency band. These plasmons are proper waves and therefore differ in sign of Im$(k_z)$ \cite{Pengchao_NP_2018}. The forward-wave branch features cutoff frequency [$\text{Re}(k_z)\approx\text{Im}(k_z)$] \cite{Pengchao_NP_2018} characterized by the reduced propagation length. For plasmons in Fig.~\ref{fig:fig8} the high-frequency and the low-frequency bands originate from spoof plasmon and MSP, respectively. It can be seen from Fig.~\ref{fig:fig8}(b) that spoof surface plasmon possesses lower losses than MSP. This can be understood more readily by comparing the dispersion properties of spoof surface plasmon with those of MSP shown in Fig.~\ref{fig:fig6}. For $u \approx 0.13$ the MSP of semiconductor nanowire coated by the axial graphene strips has the same frequency ($\approx53$~THz) and the axial wavenumber [$\text{Re}(k_z R)=0.5$] as spoof surface plasmon of Fig.~\ref{fig:fig8}. Despite the fact that the MSP’s propagation length of about 2.7~$\mu$m is relatively high in this case, the spoof surface plasmon of the semiconductor nanowire coated by graphene-based metasurface features much higher propagation length of 105~$\mu$m for the same $\omega$ and $\beta=\text{Re}(k_z)/k$. Note also that this plasmon can be tuned to the backward-wave propagation region with increase in $\theta$ when needed. Low losses, ultraslow or backward-wave dispersion suggests broad potentials for use of such plasmons in miniaturized optical devices.

\section{Conclusions}

Tensor of effective conductivity for periodic strip grating placed at the interface between two media has been generalized to the case of graphene strips. The tensor is accurate to first-order in period-to-wavelength ratio for waves propagated in arbitrary direction with respect to the strips. The resulting graphene-based metasurface was considered as a coating for cylindrical semiconductor nanowire. For such guiding structure, two independent dispersion equations describing hybrid TE-like and TM-like waves have been derived. 

Special attention has been paid to terahertz TM-like surface plasmons. Two types of such plasmons have been revealed. The surface plasmons of one type originate from the ordinary surface plasmon of a graphene-coated semiconductor nanowire and are called the modified surface plasmons (MSPs). Usually their frequencies increase with increasing gaps between graphene strips. The reason is enhanced metasurface conductivity, which depends on whether the graphene strips are axial, azimuthal, or helical. For the azimuthal graphene strips, the gaps between strips initiate steeper rise in frequency of MSPs. However, this is of little practical importance for low-loss photonic devices, because in this case the surface plasmons attenuate rapidly with frequency. By contrast, the propagation length for MSPs of the semiconductor nanowire coated by the axial graphene strips may enhance with expanding gaps between strips. As a result, such gaps can ensure several-fold increase in frequency of MSP for the same losses as for ordinary surface plasmon of a graphene-coated semiconductor nanowire. The surface plasmons of another type are spoof plasmons, which are nonexistent in the case of nanowire coated with homogeneous graphene. For these plasmons, high-frequency limit has been determined. Depending on the coil angle of graphene strips, the spoof plasmons were found to be surface or bulk. 

It has been shown that spoof surface plasmon and MSP can interact with each other. In such case, the anti-crossing effect takes place and the dispersion curves of plasmons split into three branches, including low-frequency MSP, forward and backward surface plasmons. The high-frequency band of the last two branches originates from the spoof plasmon and was found to be low-loss. This makes possible propagation of THz surface plasmons along the semiconductor nanowire with the propagation length in excess of $0.1$~mm. Moreover, the frequency, field confinement and group velocity of spoof surface plasmons can be tuned in a wide range by adjusting the width of graphene strips and their coil angle. The exceptions are plasmons with zero group velocities, which are forbidden for nonzero Ohmic losses in graphene.

\bibliography{graphene-based}
\bibliographyfullrefs{graphene-based}
 
\end{document}